\documentclass[twocolumn]{aastex63}
\DeclareMathAlphabet{\mathsc}{OT1}{cmr}{m}{sc}
\def\testbx{bx}%
\DeclareRobustCommand{\ion}[2]{%
\relax\ifmmode
\ifx\testbx\f@series
{\mathbf{#1\,\mathsc{#2}}}\else
{\mathrm{#1\,\mathsc{#2}}}\fi
\else\textup{#1\,{\mdseries\textsc{#2}}}%
\fi}
\usepackage{multirow}
\usepackage{lineno}
\received{}
\revised{}
\accepted{}

\shorttitle{Reverberation mapping of lensed quasars}
\shortauthors{Golubchik et al.}

\begin{document}


\title{Reverberation mapping of high-mass and high-redshift quasars using gravitational time delays}

\correspondingauthor{Miriam Golubchik}
\email{golubmir@post.bgu.ac.il}

\author[0000-0001-9411-3484]{Miriam Golubchik}
\affiliation{Department of Physics,
Ben-Gurion University of the Negev, P.O. Box 653,
Be'er-Sheva 84105, Israel}
\author[0000-0003-3780-6801]{Charles L. Steinhardt}
\affiliation{Cosmic Dawn Center (DAWN), Denmark}
\affiliation{Niels Bohr Institute, University of Copenhagen, Lyngbyvej 2, Copenhagen Ø 2100, Denmark}
\author[0000-0002-0350-4488]{Adi Zitrin}
\affiliation{Department of Physics,
Ben-Gurion University of the Negev, P.O. Box 653,
Be'er-Sheva 84105, Israel}
\author[0000-0002-7876-4321]{Ashish K. Meena}
\affiliation{Department of Physics,
Ben-Gurion University of the Negev, P.O. Box 653,
Be'er-Sheva 84105, Israel}
\author[0000-0001-6278-032X]{Lukas J. Furtak}
\affiliation{Department of Physics,
Ben-Gurion University of the Negev, P.O. Box 653,
Be'er-Sheva 84105, Israel}
\author[0000-0002-4830-7787]{Doron Chelouche}
\affiliation{Department of Physics, Faculty of Natural Sciences, University of Haifa, Haifa 3498838, Israel}
\affiliation{Haifa Research Center for Theoretical Physics and Astrophysics, University of Haifa, Haifa 3498838, Israel}
\author[0000-0002-9925-534X]{Shai Kaspi}
\affiliation{School of Physics \& Astronomy and the Wise Observatory, The Raymond and Beverly Sackler Faculty of Exact Sciences Tel-Aviv University, Israel}



\begin{abstract}
Mass estimates of black holes (BHs) in the centers of Active Galactic Nuclei (AGN) often rely on the radius-luminosity relation. However, this relation, usually probed by reverberation mapping (RM), is poorly constrained in the high-luminosity and high-redshift ends due to the very long expected RM lag times. Multiply imaged AGN may offer a unique opportunity to explore the radius-luminosity relation at these ends. In addition to comprising several magnified images enabling a more efficient light-curve sampling, the time delay between multiple images of strongly lensed quasars can also aid in making such RM measurements feasible on reasonable timescales: If the strong-lensing time delay is, for example, of the order of the expected RM time lag, changes in the emission lines in the leading image can be observed around the same time as the changes in the continuum in the trailing image. In this work we probe the typical time-delay distribution in galaxy-cluster lenses and estimate the number of both high-mass ($\sim10^9-10^{10}$ M$_{\odot}$), and high-redshift ($z\gtrsim4-12$) quasars that are expected to be strongly lensed by clusters. We find that up to several dozen thousand M$_{BH}\sim10^{6}$-$10^{8}$ M$_{\odot}$ broad-line AGN at $z>4$ should be multiply imaged by galaxy clusters and detectable with JWST, hundreds with \textit{Euclid} and several thousands with the \textit{Roman} Space Telescope, across the whole sky. These could supply an important calibration for the BH mass scaling in the early Universe.

\end{abstract}

\keywords{Quasars -- galaxies: clusters: general -- gravitational lensing: strong}

\section{Introduction}
\label{sec:intro}

Super massive black holes (SMBHs) are found in the centers of most galaxies. The masses of those SMBHs are of much interest as their evolution is not well understood, and since they are believed to play a major role in their host galaxy evolution. In fact, tight relations seem to hold between the mass of a SMBH and the properties of its host galaxy such as the luminosity or the velocity dispersion of the bulge \citep[e.g.,][]{FerrareseMerritt2000Scaling,KormendyHo2013,Reines2015,Ding2020SMBHhostrelation}. Measuring SMBH masses as a function of redshift can also shed light onto black hole (BH) formation and the physics of accretion processes \citep[e.g.,][]{Wang2021Quasarz764,Volonteri2012Sci...337..544V}, and their mass evolution through cosmological time might help constrain different cosmological models \citep[e.g.,][]{Fanidakis2012AGNcosmo,Habouzit2022MNRAS.511.3751H}.

SMBHs are believed to be the central engines of Active Galactic Nuclei (AGNs). The rapid accretion onto the BH results in a high luminosity across a broad range of wavelengths and unique spectral features from which one can estimate the virial mass of the BH. Using a single epoch spectrum, one can resolve the velocity of the material in the broad line region (BLR) surrounding the BH and, through further assumptions of virial motion, obtain an estimate for the BH mass \citep[e.g.,][]{GreeneHo2005}. However, this technique strongly relies on the ability to determine the distance $R$ from the central engine to the BLR (often referred to as the size of the BLR) which is usually obtained through empirical scaling relations with the AGN's continuum luminosity $L$ (often referred to as an $R$-$L$ relation).

$R$-$L$ relations can be empirically derived using Reverberation Mapping (RM). The RM technique aims to measure the distance to the BLR in AGNs by measuring the time lag between variations in the continuum flux from the accretion disk, and the response of the emission lines in the BLR to those variations, assuming that this response propagates with the speed of light \citep{Bahcall1972RM, Blandford&McKee1982RM, Peterson1993RM,Netzer&Peterson1997}.
Almost a 100 AGNs were studied using RM to determine the sizes of their BLRs, resulting in an empirical relation connecting the continuum luminosity to the BLR size for different emission lines \citep{Kaspi2000RM,Kaspi2005RM,Bentz2009RM,Bentz2013RM}. Indeed, those R-L relations are widely used for virial mass estimates of SMBHs in the centers of AGN, varying in luminosity and redshift. However, because those relations were mostly obtained using (relatively) low redshift and lower-luminosity AGN, RM studies of high-luminosity and high-redshift quasars are needed in order to extend and calibrate these relations to those extreme ends where the relation is not well established and where the physical picture may be different (for example, the BLR might not be necessarily virialized \citep[e.g.][]{King2024MNRAS.531..550K,Lupi2024AGNMass}). 

Unfortunately, RM of high-luminosity and high-redshift quasars faces several difficulties. High-luminosity quasars are presumed to have large BLR sizes, resulting in long RM time lags between continuum and emission-line variations. Similarly, high-redshift quasars and AGN \citep[e.g.,][]{Wang2021Quasarz764,Larson2023ApJ...953L..29L,Goulding2023ApJ...955L..24G,Furtak2024shadows} will also show longer observed RM time lags compared to their lower-redshift counterparts, due to cosmological time dilation. As a result, observations for several decades are needed for those RM campaigns \citep{Kaspi2018,Kaspi2021review}. Nevertheless, several attempts have been recently made to conduct RM for those types of AGN. In one study, \citet{Kaspi2021RM} observed photometrically 11 quasars at redshifts $\sim 2<z<3.4$ and luminosities $ \sim 10^{46.9}<\lambda L_{\lambda}(1350 \AA)<10^{48.0} \ erg \ s^{-1}$ for almost twenty years while six of them were monitored spectrophotometrically for 13 years. Their study manages to significantly measure RM time lags of the C\,\textsc{iv}\,$\lambda\lambda1548,1551$\AA\ doublet in three objects and a C\,\textsc{iii}]\,$\lambda\lambda1907,1909$\AA\ RM time lag in one object. Scaling the BLR size with the UV continuum luminosity they conclude that their objects lie on the previously measured relation. \citet{Shen2023SDSSRM} preformed RM using 11-year photometric and a 7-year spectroscopic data for 849 broad-line quasars. They report 339 significantly measured RM time lag detections for the different emission lines. Although they claim that on average the R-L relations follow existing ones, they report an intrinsic scatter of $\sim 0.3$ dex in the lags for the H$\beta$ line at 4861\,\AA\ and the Mg\,\textsc{ii}]\,$\lambda\lambda2796,2803$\AA\ doublet, whereas the C\,\textsc{iv} R-L relation has an even more substantial scatter of $\sim 0.5$ dex. It is thus becoming important to understand whether these large scatters stem from intrinsic physical processes at high luminosities \citep{Du2018RLrelation,Czerny2019RLrelation,Alvarez2020RLrelation} or due to the fact that there is often a difficulty to significantly measure RM time lags for such objects given the monitoring times needed.

In this work, we discuss the possibility to probe the R-L relation for the high-luminosity or high-redshift regimes using multiply imaged quasars. If the strong-lensing (SL) time delays (TD) between different multiple images of a lensed quasar are of the same order as the RM time lags, the expected changes in the emission lines can be observed in a leading multiple image around the same time as the changes in the continuum from a trailing multiple image, so that RM of these types of objects can be done with a reasonable and shorter monitoring time-span than for similar, un-lensed quasars. Although the monitoring time will be nominally shorter by one RM time lag, we also discuss here that in practice the gain will be larger, as a lensed quasar will be lensed into several magnified images, supplying a more efficient sampling of the light curve for RM. It is worth mentioning that high quality light curves are also an essential part of velocity-resolved RM \citep[e.g.,][]{Bentz2008RM,Pancoast2014VelocityResolvedRM,Du2018velocityResolvedRM}. This is a powerful technique aiming to constrain the kinematics and geometry of the BLR through measuring the RM time lags at various velocities, not only the mean RM time lag as in classical RM. That means that the signal-to-noise and cadence need to be sufficiently high, which might be applicable due to the mentioned lensing effects. In particular, as for the classical RM, at high redshifts the process becomes even more difficult, resulting in the lack of any successful velocity-resolved measurements in this regime.

About a handful of quasars and AGN, mostly around cosmic noon, are known to be multiply imaged by galaxy clusters \citep[e.g.,][]{Inada2003Natur.426..810I,Oguri2013MNRAS.429..482O,sharon2017,Acebron2022QSO,Furtak2023z2p06AGN,Napier2023ApJ...954L..38NQuasarNew}, whereas some have also been used for cosmography \citep[e.g.,][]{Napier2023HubbleConstant}. Notably, a successful RM of a strongly lensed quasar has been recently done by \citet{Williams2021RMlensedQSO} where they exploit the TD between the multiple images of the strongly lensed quasar SDSS J2222+2745 to construct a detailed light curve spending over 6 years in the observed frame with 4.5 years of observations with high cadence. They significantly detect a RM time lag of $\tau_{cen} = 36.5^{+2.9}_{-3.9}$ rest-frame days for the C\,\textsc{iv} emission line, whereas the TD from the two other images they use are of $\sim40$ and $\sim700$ days. The quasar SDSS J2222+2745 ranged in continuum luminosity with the average value of log$_{10}$[$\lambda L_{\lambda}(1350 \AA)$/erg  s$^{-1}$] = $44.66 \pm 0.18$. In this illustrative case the RM time lag was short enough so that the gain from image multiplicity was mainly manifested in additional data points on the light curve, thus importantly allowing to shorten the overall monitoring time. Here, we wish to focus on objects at the high luminosity end, namely continuum UV luminosities $L_{UV} > 10^{46.5}$ erg\ s$^{-1}$ (BH masses of log$_{10}$M$_{BH}\gtrsim9$), or at high redshifts (i.e., $z\gtrsim4$), for which we expect RM time lags of $\sim300 - \sim1000$ days, depending on redshift. For example, a redshift $z\sim7$, $M_{BH} \sim 10^{9} - 10^{9.5}$ M$_{\odot}$ quasar \citep[e.g.][]{Wang_2021} would be expected to show an observed RM time lag of about $3-6$ years, requiring decades of monitoring to measure it without lensing multiplicity. In this paper we suggest that SL TDs of lensed quasars may be a useful route to constraining the R-L relation at those regimes. This is becoming especially timely now that JWST has been uncovering a new abundant population of Type I AGN at high redshifts \citep[e.g.,][]{Kocevski2023ApJ...954L...4K,Harikane2023,Matthee2023reddots,Larson2023ApJ...953L..29L,Greene2023arXiv230905714G}, including the first example of a cluster-lensed multiply imaged one at $z\sim$7 \citep{Furtak2023z7p6AGN,Furtak2024shadows}. 

The work is constructed as follows. In section \ref{sec:dist} we start by estimating the expected RM time lags for high-luminosity and high-redshift quasars (\S \ref{sec:RM}), and then estimate the distribution of TDs between the multiple images of lensed systems in typical lensing clusters (\S \ref{sec:SLTD}). In section \ref{sec:number_of_lensed} we estimate the expected number of lensed quasars and AGN that could be found with e.g., the JWST, \textit{Euclid}, or with the \textit{Roman} Space Telescope, in the near future. The results of this calculation are given in section \ref{sec:results}. In section \ref{sec:technique} we demonstrate the suggested idea using simulated light curves and discuss its applicability, and in section \ref{sec:Summary} we conclude the work.
Throughout we assume a standard flat $\Lambda$CDM cosmology with $H_0=70\,\frac{\mathrm{km}}{\mathrm{s}\,\mathrm{Mpc}}$, $\Omega_{\Lambda}=0.7$, and $\Omega_\mathrm{m}=0.3$. Magnitudes are quoted in the AB system \citep{Oke1983ABandStandards} and all quoted uncertainties represent $1\sigma$ ranges unless otherwise stated. 

In this work we use the terms ``quasars" and ``AGN" synonymously and whenever we discuss lensed AGN in this work, we explicitly refer to multiply imaged examples. In addition, in this work “time delay” always refers to the gravitational lensing time delay, and for RM time lag we always use the term “time lag”.

\section{The distributions of RM time lags and lensing time delays}\label{sec:dist}
\subsection{RM of quasars: the distributions of RM time lags}
\label{sec:RM}

\begin{figure}
    \centering
    \includegraphics[width=\linewidth,trim={0.8cm 0.4cm 1.5cm 1.6cm}, clip]{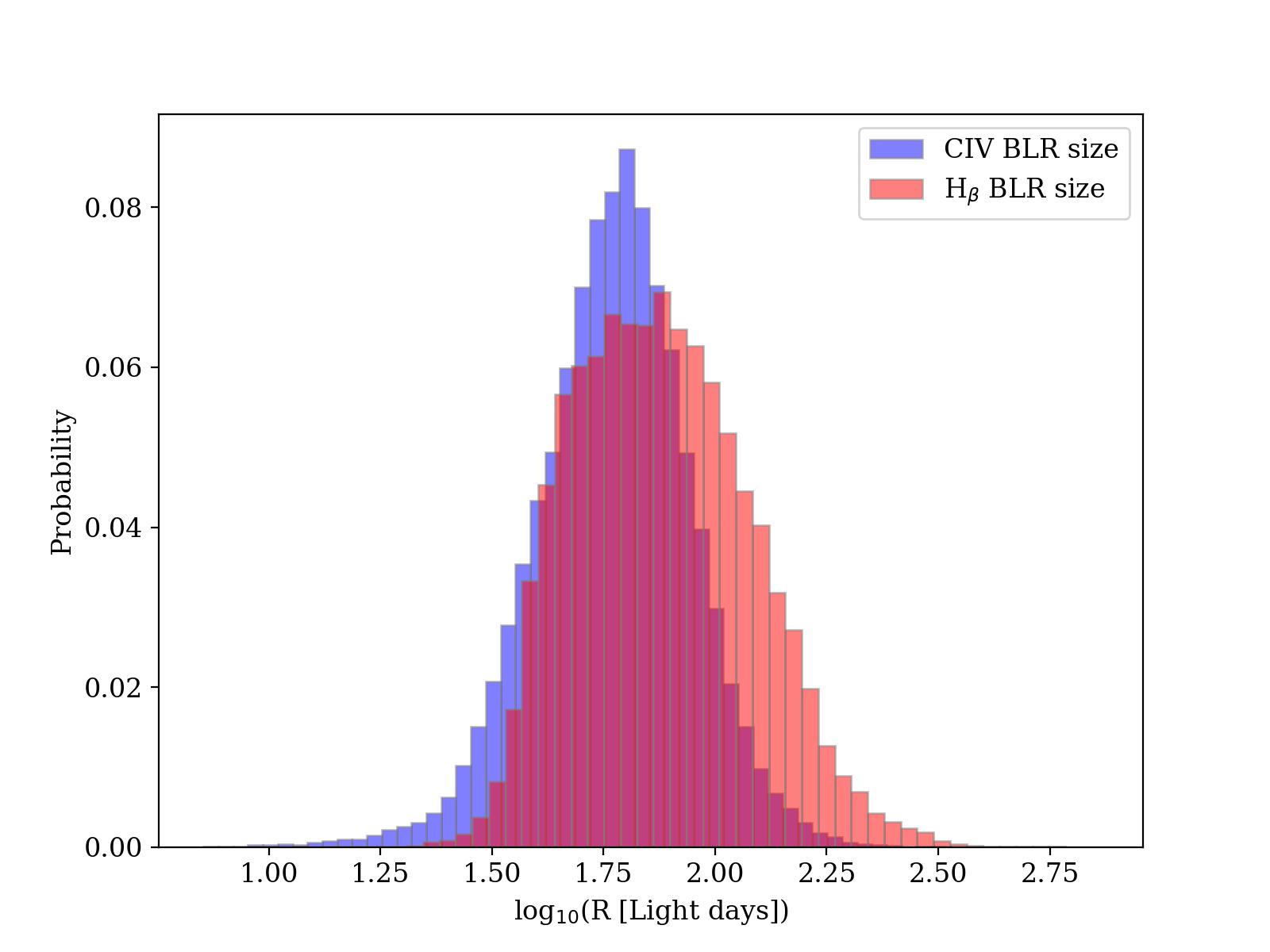}
    \caption{Histograms of the expected BLR sizes for SDSS DR7 quasars. The catalog contains two distinct subsamples (see sub-section \ref{sec:RM} for details). The H$\beta$ RM time-lag distribution is plotted for the subsample of quasars which have an optical continuum measurement, assuming the R-L relation by \citet{Bentz2013RM}. The C\,\textsc{iv} time-lag distribution is plotted for the subsample of quasars which have a UV continuum measurement, using the relation from \citet[][]{Kaspi2021RM}. The distributions seem to each fit well to a Gaussian distribution, with $[\mu,\sigma]$=$[1.86,0.21]$ and $[\mu,\sigma]$=$[1.78,0.16]$ for H$\beta$ and C\,\textsc{iv}, respectively. For example, $\sim9$\%($\sim26$\%) of the expected C\,\textsc{iv}~(H$\beta$) rest-frame RM time lags will be $100$ days or longer. This means $500-1300$ observed days, depending on redshift. The reader should also note that in general, \emph{for the same quasar}, H$\beta$ RM time-lags are expected to be longer by a factor of a few compared to C\,\textsc{iv} RM time-lags, so that in practice the H$\beta$ distribution may extend to even longer RM time lags.}
    \label{fig:lagHbeta}
\end{figure}

\begin{figure}
    \centering
    \includegraphics[width=\linewidth,trim={0cm 0.2cm 1.0cm 1.0cm}, clip]{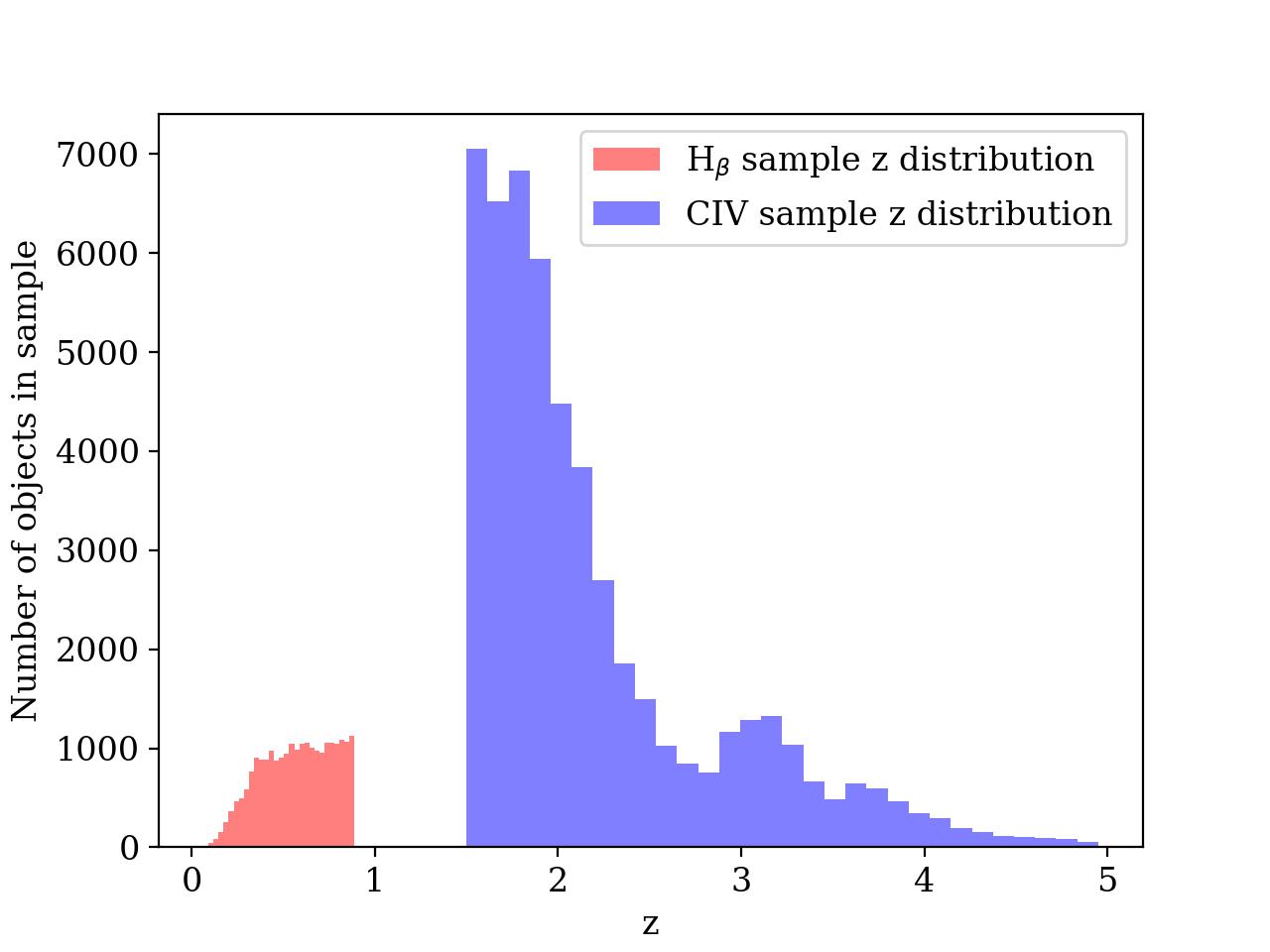}
    \caption{Histogram of the redshift of the two subsamples of SDSS DR7 quasars used to construct the distributions in Fig.\ref{fig:lagHbeta}. The AGN constructing the $H_{\beta}$ RM time-lag distribution
    are all at redshift $<1$ with a median redshift of 0.59.
    As for the quasars used to construct the CIV RM time-lag distribution,  $80 \%$ are between redshifts $1.5$ and $2.5$, with a median redshift of 1.9.}
    \label{fig:z_dist}
\end{figure}

As was noted above, RM of quasars relies on the ability to detect temporal variations in the continuum emission from a quasar that lead to, after a time lag which depends on the BLR size, similar variations in the emission lines from the BLR. In this section, we estimate the BLR sizes -- and thus RM time lags -- of a representative sample of quasars, namely the SDSS quasar catalog, which is based on Data Release 7 (DR7) \citep{ShenYueSDSSCatQuasars2011} and is available online. This catalog supplies continuum luminosities for either $\lambda L_{\lambda}(1350 \AA)$ or $\lambda L_{\lambda}(5100 \AA)$, depending on redshift. 

The distributions of BLR sizes are shown in Fig.\ref{fig:lagHbeta}. The expected H$\beta$ RM time-lag distribution is obtained assuming the R-L relation from \citet{Bentz2013RM}, for those quasars that have a measurement of the continuum luminosity at 5100 \AA.
The expected C\,\textsc{iv} RM time-lag distribution is obtained assuming the R-L relation from \citet{Kaspi2021RM} (see their eq.~1), using the continuum UV luminosity at 1350 \AA.  Note that the C\,\textsc{iv} subsample consists of 52,427 objects, which have a higher redshift (median redshift of 1.96) and are typically more luminous (median $\lambda L_{\lambda}(1350 $\AA$)=10^{46.1}$ erg s$^{-1}$) than the H$\beta$ subsample (23,093 objects, median redshift of 0.59, median $\lambda L_{\lambda}(5100 $\AA$)=10^{44.6}$ erg s$^{-1}$), the redshift distribution of both subsamples is shown in Fig.\ref{fig:z_dist}.
As can be seen in Fig. \ref{fig:lagHbeta}, the BLR size distributions, plotted as log$_{10}$(R) where R is in light days, are approximately Gaussian, with a mean and width of $[\mu,\sigma]$=$[1.86,0.21]$ and $[\mu,\sigma]$=$[1.78,0.16]$ for H$\beta$ and C\,\textsc{iv}, respectively. About $9$\%, roughly, of the quasars show C\,\textsc{iv} RM time lags of $100$ days or longer in the rest-frame, and about $26$\% of the quasars show H$\beta$ RM time lags of $100$ days or longer in the rest-frame. This translates to $500-1300$ days for redshifts of $z\sim4-12$, respectively, which we define here as the high redshift regime. 
Since we explore the typical RM time lags in the rest frame, we should note the intrinsic redshift distribution of the sample. For $H_{\beta}$ all AGN are at redshift $<1$ (i.e. in a relatively narrow redshift range, see Fig. \ref{fig:z_dist}).
For CIV,  $80 \%$ are between redshifts $1.5$ and $2.5$, and the rest are in the tail (extending to redshift $z\sim5$). When we construct the CIV BLR size distribution from only those $80 \%$ that are at $z \sim2$ , we get a similar shape distribution with the mean value shifted slightly to lower values (by $2 \%$). AGN at redshift, for example, $4-5$ constitute $2 \%$ of the total population used to create the CIV BLR size distribution and when using only them, the shape of the distribution still holds, although the mean value of the BLR size distribution is shifted to higher values by $11 \%$. This shift is most likely due to the fact that at higher redshifts, for a given observational depth, only more luminous AGN could be observed. These deviations are thus small enough, allowing us to assume in the following subsection \ref{sec:SLTD} and upcoming Fig.\ref{fig:overlap} that the distributions in Fig.\ref{fig:lagHbeta} are representative for a population of AGN at any redshift bin. 
Note that this procedure does not take into account a possible evolution of the BLR size with redshift (which may not be very significant e.g., \citealt{Du2018velocityResolvedRM}), nor other observational biases.

\subsection{The distribution of strong-lensing time delays}
\label{sec:SLTD}
In this subsection we aim to obtain a general distribution of the TDs caused by strong lensing in clusters. Galaxy clusters act as gravitational lenses, causing the bending of light. When the projected mass density is larger than some critical density for lensing, strong lensing will occur comprising multiple images of background sources. Using the thin lens approximation, light that travels from a source to the observer will be delayed by some time $t$ \citep[][]{Schneider1985}:
\begin{equation}
    \centering
    \label{eq1}
    t(\vec{\beta},\vec{\theta}) = \frac{D_{l}D_{s}}{D_{ls}}\left[\frac{1}{2}(\vec{\theta} - \vec{\beta})^2 - \psi(\vec{\theta})\right]\frac{(1+z_{l})}{c} ,
\end{equation}
with $D_l$, $D_s$ and $D_{ls}$ being the angular diameter distances to the lens, source and between the lens and the source, respectively; $z_l$ is the redshift of the cluster; $\vec{\theta}$ and $\vec{\beta}$ represent the position of the image and source, respectively; and $\psi(\vec{\theta})$ is the lensing potential in that position in the image plane (see \citet{NarayanBartelmann1996Lectures} for more details).
The relevant lensing quantities can be obtained for different clusters using SL lens modeling \citep[e.g.,][]{Richard2014FF,Mahler2019MACS0417,bergamini22}. We use mass models made previously by our group \citep[e.g.,][]{Zitrin2014CLASH25,Furtak2022UNCOVERlensmodel,Furtak2023z7p6AGN,Furtak2023z2p06AGN} for three typical high Einstein mass galaxy clusters: RXC0018, MACS0035 and Abell 2744, where the Einstein mass is the projected mass enclosed within the Einstein radius, and three typical lower Einstein mass clusters: Abell~383, Abell~611 and RX~J2129. While all these clusters have typical masses of a few times $10^{14}$ M$_{\odot}$ to $\sim10^{15}$ M$_{\odot}$, the high Einstein mass clusters have a corresponding Einstein radii of $\sim20-30\arcsec$ (and thus higher enclosed mass) and the low Einstein mass clusters have Einstein radii of about $\sim10-15\arcsec$ typically. Two of the three high mass clusters also happen to show a multiply lensed AGN or quasar-like system with three multiple images each: MACS0035 contains a Type~I AGN at $z=2.06$ \citep{Furtak2023z2p06AGN}, and a redshift $z=7.05$ ``red-dot" AGN was found behind Abell~2744 \citep{Furtak2024shadows,Furtak2023z7p6AGN}.

\begin{figure*}
    \centering
\includegraphics[width=\linewidth,trim={0.1cm 0.2cm 0.5cm 0.8cm}, clip]
{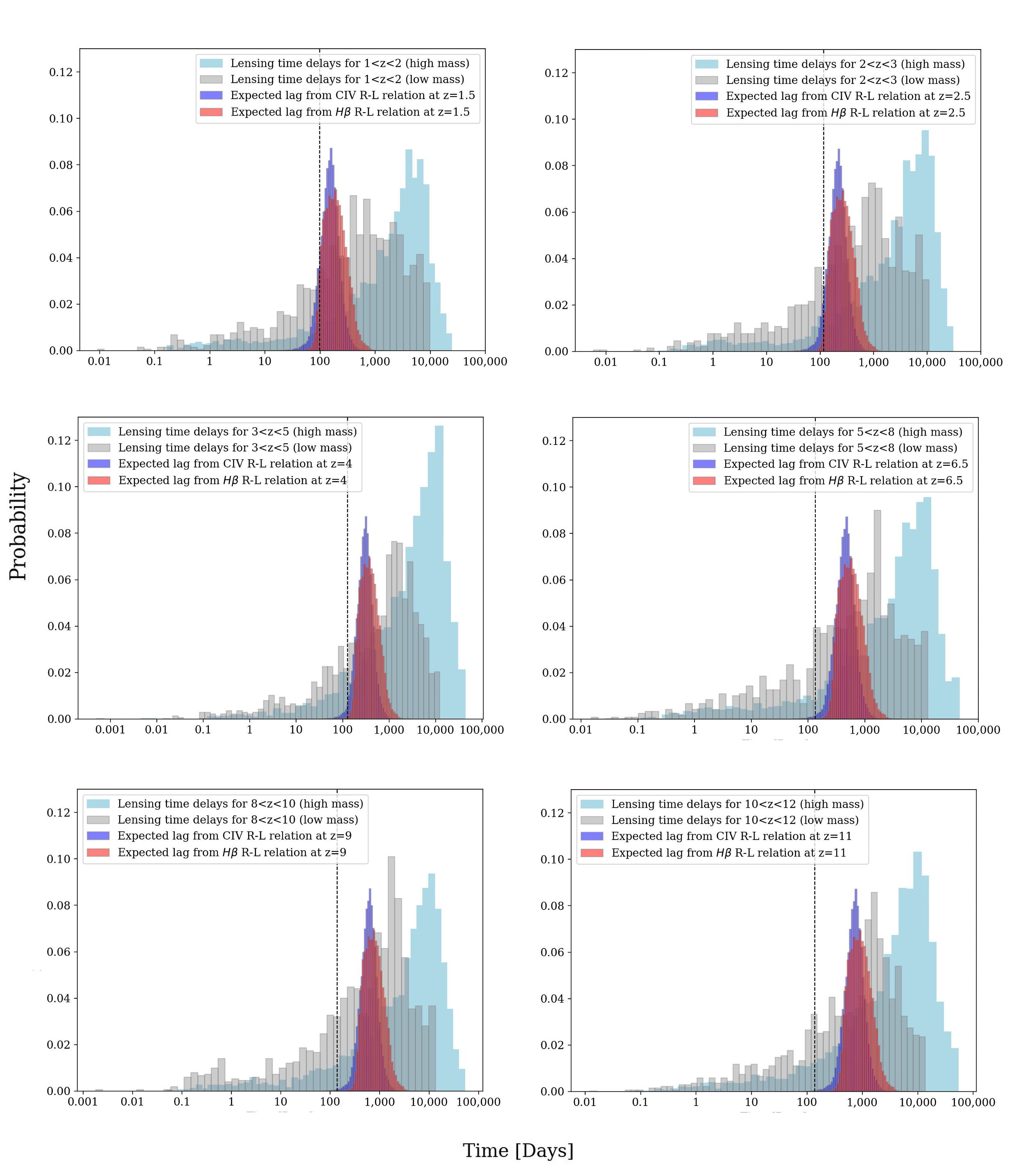}
\caption{Overlap of the distribution of SL TDs by lensing clusters, and the distribution of expected RM time lags from the BLR sizes for different redshifts. We show the distribution for both high and low mass galaxy clusters. The redshift limits in each panel's legend indicate the redshift range of the sources that are lensed to create the SL TD distribution. The RM time lags are corrected for the specified, central redshift, i.e., they are given in the observed frame as well. \emph{As can be seen, there is substantial overlap such that SL TDs by galaxy clusters completely cover the range of expected RM lag times.} We also mark with a dashed black line a typical TD caused by galaxy-galaxy strong lensing by a massive galaxy (where we adopt a SIS galaxy model with $\sigma=250$ km s$^{-1}$ placed at $z=0.5$), showing that only galaxy clusters supply large enough TDs for our purpose. As can be seen, smaller galaxy cluster lenses show more overlap, but they are also expected to show less multiple images per cluster. For more details see section \ref{sec:dist}.}
\label{fig:overlap}
\end{figure*}

In order to create the TD distribution, we randomly plant sources in the source plane and re-lens them to the image plane using the deflection fields from the model. Typically we use a 100 multiply imaged sources in each case. We calculate the TDs between all available pairs using eq. \ref{eq1} and repeat this process for different source redshifts, in the ranges of $1<z<2$, $2<z<3$, $3<z<5$, $5<z<8$, $8<z<10$ and $8<z<12$. The SL time-delay distributions are plotted in Fig.\ref{fig:overlap}.

We parameterize the TD distributions and note that they can be well fit (R-square of $\simeq0.98$) by the following parametrization, inspired by a log-normal distribution function:,  $P(x)= \frac{1}{a\cdot(d-x)}\cdot\exp\{{-\frac{[\ln(d-x)-b]^2}{c^2}}\}$, with $a\sim10-16$ typically (depending on the redshift bin), $b\sim0.1$, $c\sim0.8$ and $d\sim4.6$ for most redshift bins. Various previous works \citep[e.g.,][]{Oguri2002TDdist,Oguri2003TDdist, Li2012TDstatistics} have also looked into the TD statistics. Another useful way to parameterize the TD distribution is in log-log space, where we find a slope of $\sim0.8$, similar to what was found for, e.g., Abell 1689 \citep{Li2012TDstatistics}.

In Fig.~\ref{fig:overlap} we also compare the lensing TD probability distributions to the RM time lags expected from the BLR sizes, recalculated to match each redshift bin ,as we assume that this distribution is representative at all redshifts, which is a sufficient approximation for our purpose, namely showing that the RM time lag distribution is of the order of TDs by galaxy clusters. More on this assumption in previous subsection \ref{sec:RM}. We see that there is an overlap in the distributions such that TDs by both types of galaxy clusters cover the entire range of expected RM time lags, with a higher probability of having the relevant TDs, i.e., of order the RM time lag, for higher-luminosity and higher-redshift quasars. About $20\% - 40\%$ of the TD pairs in all clusters overlap with the range of RM time lags, depending on the redshift of the lensed source. We see that the peak of the TD distribution caused by lower mass clusters, which should be more abundant, agrees better with the peak of the RM time lag distributions, but those clusters would have also less multiple images per cluster in proportion to their smaller critical area size (which is roughly proportional to the square root of the mass), such that both populations should be useful.
Also overplotted is a line marking a typical TD expected by a massive galaxy lens -- where we nominally adopt for this calculation a SIS galaxy of $\sigma_{vel}\sim250$ km s$^{-1}$ placed at redshift 0.5 \citep{Oguri2019transients}, demonstrating that TD in the case of lens galaxies are too short for a significant gain in the wait time needed for RM of high-redshift, high-mass quasars, and that galaxy clusters are needed for that purpose.  

\section{Expected numbers of lensed quasars}\label{sec:number_of_lensed}
To examine whether the proposed idea for RM of multiply imaged quasars at the high luminosity and high redshift ends might become realistic in the near future, we estimate the expected number of quasars that would be multiply imaged by galaxy clusters. Given the many uncertainties involved (see for example subsection \ref{subsec:obs_diff}), our goal is merely to obtain an order-of-magnitude estimate. We focus on the observational properties of three observatories: \textit{Euclid}, the \textit{Roman} Space Telescope, and JWST. 
\textit{Euclid} is an optical and NIR space telescope with a mirror of size 1.2 meters in diameter, covering the wavelengths between 550 nm and 2 $\mu$m. Covering the same wavelength range, the soon to be launched \textit{Roman} Space Telescope will have a 2.4 m primary mirror and will achieve greater depth with the same observation time. Lastly, JWST is mainly a NIR space telescope with a 6.5m mirror, with wavelength coverage ranging from 0.6 $\mu$m to 28.3 $\mu$m. 
The basic idea is to calculate for each observatory, i.e., a filter set and depth limit, how many quasars in the relevant BH mass and redshift regimes fall within the caustics of lensing clusters so that they would be multiply imaged, and would be bright enough, folding in the magnification, to be detected. We adopt relevant quasar luminosity functions (QLFs) and generally follow known prescriptions for the lensing optical depth, where we assume an explicit cluster mass function -- from which an Einstein radius function is calculated -- and then calculate the magnification bias and expected fraction (and total number) of multiply imaged quasars in the relevant cluster mass and redshift ranges. 

\subsection{The Quasar Luminosity Function}\label{subsec:QLF}
For the calculations below we adopt two complementary QLFs: For UV selected quasars we use the one from \citet{Willott2010QLF}, and to account for recent populations of AGN uncovered with JWST for example, we also use the QLF from \citet{Harikane2023}. Both QLFs are given in terms of the UV absolute magnitude at 1450\,\AA, $M_{1450}$. The former also has a term for evolution with redshift, which we assume applies for the range of redshifts spanned here ($4<z<12$), while the latter is divided into four discrete QLFs, one for each of the following redshift bins - z$\sim4$, z$\sim5$, z$\sim6$, and z$\sim7$, where we assume that it does not change significantly from z$\sim7-12$ -- but as seen below, even as such the redshift window from z$\sim8-12$ only negligibly contributes to the result.
Note that, given that we are interested in an order of magnitude and since the results heavily rely on the various assumptions made throughout, for the QLFs we adopt only the best-fit parameters, without propagating their uncertainties (which would affect the uncertainty on our estimates, too). Also, while the QLF from \citet{Willott2010QLF} is less constrained at the faint end, the QLF from \citet{Harikane2023} is better constrained at that regime.

\subsection{The total expected number of quasars}\label{subsec:first_estimates}
For each observatory we then do the following. We first divide the relevant redshift range into bins. For each bin, we redshift a quasar's SED and define the selection according to the band in which the quasar would be brightest. 
Then, using the QLF, which itself is divided into magnitude bins for summation, we calculate how many quasars are expected in that redshift bin down to the observational depth. The \emph{total} number of quasars expected to be observable across the sky is then obtained by summing this number up over the different redshift bins. 

For the \textit{Euclid} survey we focus on the detection of typical "blue" quasars and adopt the spectrum from \citet{VandenBerk2001}. Following the \textit{Euclid} overview paper \citep{Euclid2011overview}, we primarily use the QLF from \citet{Willott2010QLF}, but also for comparison employ the \citet{Harikane2023} one. The observational depth for the different bands is also adopted from \citealt{Euclid2011overview}. For each redshift bin we choose as the selection band the band in which the rest frame $1450$\AA \ falls. If the redshifted wavelength is lower (or higher) than the survey's available bands, we choose the nearest band to it. Since in this work we are particularly interested in lensed high-mass and high-redshift quasars, we perform the calculation for those two regimes specifically. 


We follow the same procedure for the \textit{Nancy Grace Roman Space Telescope}. Both observatories cover roughly similar wavelength ranges into the near-infrared. While the nominal FOV for \textit{Roman} is about half of that of \textit{Euclid}, \textit{Roman} has a two times in diameter larger mirror and the observational limits are deeper than the ones for \textit{Euclid}, hence we may expect higher numbers of quasars to be detectable with it. We adopt the depth values for the different bands from the \textit{Roman} Space Telescope website \footnote{\url{https://www.stsci.edu/files/live/sites/www/files/home/roman/_documents/roman-science-sheet.pdf}}. Again our focus remains on the high-redshift and high-luminosity quasars and as above we adopt the typical SED from \citet{VandenBerk2001}.


Last we turn to perform the same calculation for JWST. In addition to typical "blue" quasars that are detectable with JWST at redshifts higher than $\sim 4$, given the new population of "red-dot" AGN at high redshift that has been discovered in JWST observations in increasing numbers \citep[e.g.,][]{Furtak2023z7p6AGN,Kokorev2023ApJ...957L...7K, Matthee2023reddots,Labbe2023RedDots,Kocevski2023Reddots,Kocevski2024reddots}, we also consider here an additional quasar SED model. The SED of this type of object is heavily reddened and thus is brighter towards the red, with some UV light from star formation or scattered AGN radiation. As the representative SED for red-dot objects we use the one from \citet{Furtak2023z7p6AGN}, and we only consider here the QLF from \citet{Harikane2023} which was based on or commensurate with this new population of quasars \citep{Greene2023arXiv230905714G}. Given the red SED, for red-dot AGNs we now choose the selection band to be the one in which the rest frame $5500$\AA~falls, shifting all selection to the red-optical end of the spectrum, which indeed, is currently only observable with JWST for these redshifts. We again focus on the high-redshift population for our calculations and estimate the expected numbers of quasars to be detected with JWST for both blue quasars and red quasars. To account for JWSTs extreme depth capabilities we assume a depth of 30 magnitudes, reachable in several hours with JWST. 


\subsection{The lensed fraction}\label{sec:detailedcalc}
Various works have calculated the effect of lensing on the luminosity function and the fraction of some population of lensed sources of interest such as galaxies, quasars or supernovae, and the prescriptions for doing so are well established. Here we largely follow the outlined procedure in \citet{Yue2022ApJ...925..169Y}, where they obtained an estimate for the fraction of lensed quasars mostly by galaxies. i.e. in a different mass regime that is our interest here, and expand it to massive galaxy clusters. Various components of the formalism are also based on e.g., \citet{OguriMarshall2010,Wyithe2011Natur.469..181W,Mason2015}. One other important difference is that in this formalism, given it was usually applied to galaxies, a singular isothermal spheres (SIS) mass density profile is typically assumed, where here we consider the Navarro–Frenk–White (NFW) mass density profile \citet{Navarro1996}, which is a more realistic approximation for the mass distribution of clusters.

One can define the optical depth for multiple images, which describes how likely it is for a source at $z_{s}$ to lie within the caustics of a lens (the Einstein radius mapped back to the source plane), and thus be multiply imaged.  The optical depth for multiple images is then given by:
\begin{equation}\label{eq:tauNFW}
    \tau_m=f_{sp}\int_0^{z_s}dz_l \int dM \phi(M, z_l) \frac{d^2V_c}{d\Omega dz_l} \pi \theta_E(M, z_l, z_s)^2 D_l^2 ,
\end{equation}
where $\phi(M,z{l})$ is the lens density function and taken here to simply be the halo mass function (see eq. \ref{eq:massFunc}), $z_l$ is the redshift of the lens, $D_l$ is the angular diameter distance to it, and $f_{sp}$ is the fraction of the critical area ($\pi\theta_{E}^{2}$) in which sources would be multiply imaged (for example, for a SIS lens, $f_{sp}=1$).
The expression $\frac{d^2V_c}{d\Omega dz_d}=(1+z_d)^3c\frac{dt}{dz_d}$ is the differential co-moving volume\footnote{Note that in some work the differential co-moving volume is defined such that it contains the $D_{l}^2$ factor seen in eq. \ref{eq:tauNFW}.}, with $\frac{dt}{dz_d} = \frac{1}{H(z)(1+z)}$ where $H(z) = H_0(\Omega_m(1+z)^3+\Omega_\Lambda)^{1/2}$. 

For our calculations we adopt the mass function from \citet{Tinker2008MF}, namely:
\begin{equation}\label{eq:massFunc}
\frac{dn}{dM} = f(\sigma)\frac{\rho_m}{M}\frac{d\ln\sigma^{-1}}{dM}
\end{equation}
with $\rho_m = \rho_{crit}\Omega_m$ being the mean matter density, $\sigma$ is the RMS variance of a spherical top hat containing a mass M and $f(\sigma)$ is the halo multiplicity function given also in \citet{Tinker2008MF}. 
In practice, to calculate the mass function (and the power spectrum required therein) we use \texttt{cluster\_toolkit} and \texttt{CLASS}. We refer the reader to \url{https://cluster-toolkit.readthedocs.io/en/latest/source/massfunction.html} for more information.
The Einstein radius of a NFW lens depends on the mass, concentration and the redshifts of lens and the source. To obtain the Einstein radius for the NFW lens for each combination of parameters we follow the procedure from \citet{BroadhurstBarkana2008} and \citet[][see also \citealt{Bartelmann1996A&A...313..697B}]{SadehRephaeli2008}, while we assume the concentration-mass (c-M) relation from \citet{Meneghetti2014CLASHsim}:
\begin{equation}\label{eq:cMrelation}
c(M,z) = A \left(\frac{1.34}{1+z}\right)^B \left(\frac{M}{8\times10^{14}h^{-1}M_\odot}\right)^C
\end{equation}

with $A=3.757\pm0.054$, $B=0.288\pm0.077$ and $C=-0.058\pm0.017$ (although note, we only use here the best fit values to translate the mass to Einstein radius; we do not propagate the scatter in this relation).

Analytically, for a NFW lens multiple images will occur if the source lies within the radius of the radial critical curve \citep{Meneghetti2003}. We however adopt here the more empirical assumption of $f_{sp}=0.1$, which we obtained from an ensemble of cluster models for known lensing clusters \citep[e.g.,][]{Zitrin2014CLASH25}: We find that in practice, the angular source-plane area enclosed by the caustics is about $\sim10\%$ of the angular area enclosed by the critical curves in the image plane (see also \citet{Broadhurst2005a}).

Next we need to consider that in flux-limited surveys, objects that lie intrinsically below the detection threshold can be detected if they are sufficiently magnified. However, there exists a trade-off with the smaller area probed in the source plane and the actual gain in number count would depend also on the shape of the luminosity function \citep{Broadhurst1995MagBias,Wyithe2011Natur.469..181W,FialkovLoeb2015ApJ...806..256F}. The magnification bias can be written as:
\begin{equation}\label{eq:Bias}
    B=\frac{\int_{\mu_{min}}^{+\infty}d\mu~p(\mu)N(>L_{lim}/\mu)}{N(>L_{lim})},
\end{equation}
where $p(\mu)$ is the magnification distribution of lensed sources, $L_{lim}$ is the survey flux limit, and $N(>L_{lim})$ is the cumulative luminosity function, i.e., the number of background sources more luminous than $L_{lim}$.

We use $\mu_{min}=2$ for the magnification
bias calculation, and, as the magnification probability function we adopt $p(\mu)\propto 1/(\mu)^3$ (truncated at $\mu=100$). This distribution seems to approximately describe the magnification distribution of lensed images, also in NFW lenses (see for example Figure 9 in \citet{Wyithe2001NFW}). We also verify this by simulating a NFW lens and calculating the magnification distribution of randomly planted sources. 

As is also indicated in \citet{Yue2022ApJ...925..169Y},
out of all sources with apparent luminosity greater than $L_{lim}$, the lensed fraction is given by \citep{Mason2015}:
\begin{equation} \label{eq:fmulti}
    F_{multi} = \frac{B \tau_m}{B \tau_m + B'(1-\tau_m)}
\end{equation}
where $B'$ is the magnification bias of non lensed sources, and we assume here it is $B'\approx1$ as in \citet{Yue2022ApJ...925..169Y,Mason2015Bias,Wyithe2011Natur.469..181W}.

\begin{deluxetable*}{lccccc}
\tabletypesize{\footnotesize}
\tablecaption{The estimated numbers of \textit{lensed} quasars that might be detected with \textit{Euclid}, \textit{Roman} space telescope and JWST.}
\label{table:TableAll}
\tablecolumns{6}
\tablewidth{0.1\linewidth}
\tablehead{
\colhead{QLF reference 
} &
\colhead{all sources
} &
\colhead{high L
} &
\colhead{high z 
} &
\colhead{C\textsc{iv} $>$ 50 days
} &
\colhead{C\textsc{iv} $>$ 100 days
} }
\tiny\startdata
\multicolumn{6}{c}{\textit{Euclid}}\\
\hline
\cite{Willott2010QLF} & $50 \ (1.3 \times 10^7)$ & $<1 \ (19,800)$ & $3 \ (84,000) $ & $3 \ (81,500)$ & $2 \ (41,800)$\\
\hline
\citet{Harikane2023} & $2,000 \ (1.5\times10^8) $ & $<1 (<1)$ & ~~~$400\ (2\times10^6)~~~ $ & $376 
 \ (1.7\times10^6)$ &~~~$36 \ (45,000)$~~~ \\ 
\hline
\multicolumn{6}{c}{\textit{Roman} Space Telescope }\\
\hline
\citet{Willott2010QLF} & $200 \ (5.9\times10^7) $ & $<1 \ (19,800)$ & $13 \ (4.2\times10^5) $ & $4 \ (1.2\times10^5)$ & $2 \ (46,000)$\\
\hline
\citet{Harikane2023} & $32,700 \ (2.4\times10^9) $ & $<1 (<1)$ & ~~~$9,800\ (1.3\times10^8)~~~ $ & $2,150 \ (1.7\times10^7)$&~~~$207 \ (1.1\times10^6)$~~~ \\ 
\hline
\multicolumn{6}{c}{JWST}\\
\hline
\citet{Harikane2023} & $1.3\times10^5 \ (8.8\times10^9) $ & $<1 \ (<1)$ & $5.3\times10^4 \ (7.2\times10^8) $ & $2,750 \ (2.4\times10^7)$ & $140 \ (1.1\times10^6)$\\
\hline
\citet{Harikane2023} (red selection) & --- & $<1 (<1)$ & ~~~$3.5\times10^4\ (5.3\times10^8)~~~ $ & $3,500 \ (2.4\times10^7)$ &~~~$190 \ (1.1\times 10^6)$~~~\\
\hline
\enddata
\tablecomments{
\footnotesize
 In every column, we also specify within the brackets the total number of quasars expected to be detected;  \emph{Column 1:} The assumed QLF. The detection limit is calculated assuming a blue selection unless otherwise specified; \emph{Column 2:} The estimated number of \textit{lensed} quasars at all redshifts and all luminosity range; \emph{Column 3:} The estimated number of \textit{lensed} quasars at all redshifts and high luminosity range; \emph{Column 4:} The estimated number of \textit{lensed} quasars at high redshifts; \emph{Columns 5-6:} The estimated number of \textit{lensed} quasars with luminosities and redshift that would result in a C\,\textsc{iv} observed RM time lag of above 50 and 100 days;}
\end{deluxetable*}


                            

After calculating the expected number of quasars to be detected as explained in the previous section, the number is multiplied by the lensed fraction $F_{multi}$ for each redshift bin. Eventually, a total lensed fraction is calculated. We perform the calculation for three different regimes - quasars at all redshifts and luminosity ranges, high luminosity quasars (i.e. $M_{1450} < -26.5$ which corresponds to $\lambda L_{UV,1450} > 10^{46.5}$ erg s$^{-1}$) and high redshift quasars, namely $4<z<12$. Last, out of the high redshift candidates we aim to obtain the number of quasars for which the notions we highlight here might be particularly useful, i.e, those with long lag times. We follow the same procedure, only this time for each redshift bin we find the lower limit for the luminosity such that the expected RM time lag of the C\,\textsc{iv} line is at least, e.g., a 100 days. As can be seen in Fig.\ref{fig:lagHbeta}, the expected RM time lags for the $H_\beta$ emission line will be even longer.

%

\section{Results}\label{sec:results}
We summarize the results in Table \ref{table:TableAll}.
For all observatories, we see that the chances to detect a multiply imaged, high-luminosity quasar are very small - with $<1$ that are expected to be detected across the sky, with both QLFs. Nonetheless, for the high redshift regime, we obtain that many such lensed AGN should be found. Explicitly, we find that $\sim 3-400$ multiply imaged AGN may be detected with \textit{Euclid} over the whole sky, with the exact number depending on the choice of QLF. For the \textit{Roman} Space telescope, we similarly estimate $\sim 13-9,800$ detectable, multiply imaged quasars, and for JWST we find that $\sim 3 \times 10^4 - 5 \times 10^4$ quasars multiply imaged by galaxy clusters might be detected all over the sky. In general, then, the numbers of multiply imaged AGN above correspond to about $\sim10^{-4}$ of the total (unlensed) population.

It is interesting to note that JWST has imaged to date about $\sim10$ clusters and in one of them, Abell 2744, a conspicuous multiply imaged AGN at $z\simeq7$ has already been discovered \citep{Furtak2023z7p6AGN}. If we take this 1 in 10 ratio as representative, then many more such multiply imaged AGN should be detected soon with JWST, as it expands its observations to a larger cluster sample. Moreover, with $\sim100,000$ clusters in the sky one could expect about $\sim10,000$ multiply-imaged AGN detectable with JWST, which is of the same order of magnitude as the result obtained in our more detailed calculation, implying that the expected number is representative. 

While it is encouraging to see that many AGN multiply imaged by galaxy clusters could be detected across the sky, it is also insightful to examine which fraction of this lensed population has a large expected RM time lag, such that the gain from lensing would be most significant. Columns 5 and 6 in Table \ref{table:TableAll} list the expected number of multiply imaged AGN that are expected to have lag times of, as an example, $>50$ and $>100$ days. For \emph{Euclid}, we expect $\sim3 - 380$  multiply imaged AGN with $>50$ days observed RM time lags and $\sim2 - 35$ with $>100$ days; $\sim4-2150$ and $\sim2-207$ with the \textit{Roman} space telescope; and $\sim2750 - 3500$ and $\sim140 - 190$ with JWST, respectively, across the sky. This is about 1-2 orders of magnitude lower than the total number of high-redshift AGN that are multiply imaged by galaxy clusters. All in all, our calculations show that there exists a large lensed population in the sky which can gain substantially from RM campaigns to probe the R-L relation at high-redshift. This may be particularly beneficial for those with longer RM lag times, as we also discuss further below (section \ref{sec:technique}).

In addition to the above, it is seen in Table \ref{table:TableAll} that for both \textit{Euclid} and the \textit{Roman} space telescope higher numbers of quasars are generally expected assuming the QLF from \citet{Harikane2023} compared to the QLF from \citet{Willott2010QLF}. The order of magnitude difference is not surprising given the high number density of broad-line AGN recently detected with JWST to fainter magnitudes \citep[e.g.,][]{Furtak2023z7p6AGN,Labbe2023RedDots,Greene2023arXiv230905714G,Kocevski2023ApJ...954L...4K,Matthee2023reddots}.

In Table \ref{table:TableAll}, for completeness, the first column lists the numbers of quasars \emph{across all redshifts} that are expected to be multiply imaged by galaxy clusters. It should be noted that both the QLFs we adopt for this calculation are, however, limited in redshift. For the \citet{Harikane2023} QLF, which is defined between redshifts $z\sim4$ and $z\sim8$, we assume no additional evolution: We use the $z\sim8$ QLF for all redshifts above $z\sim8$, which contributes only negligibly to the final numbers, and the $z\sim4$ QLF to all redshift below $z\sim4$. Since the QLF is expected to rise down to $z\sim2$ and then turn over in resemblance to the cosmic star formation density \citep[e.g.,][]{Richards2006QLFevolutionZ,SHENQLFevolutionwithZ2020}, we note that the exact effect of this assumption is not clear, in particular since the evolution of `` little red-dot" and other JWST selected AGN is not well established yet to lower redshifts. Regarding the QLF from  \citet{Willott2010QLF}, we note that it has a term for a monotonic evolution with redshift, although it was defined in practice over the redshift range $\sim 5.5-6.5$ and the redshift evolution term assumes a monotonically increasing number density towards lower redshifts. In that case, assuming this QLF below $z \sim 4$ might cause an over estimate of the number of quasars at the low redshift regime. A detailed examination of the effect of the $z\sim1-4$ range on the total numbers is out of scope for this work, in particular, since here we focus on the high-redshift regime $z \gtrsim4$.

\section{Discussion}\label{sec:technique}

To examine the feasibility of the idea proposed in this work, in the first part of this work we have shown the distribution of TDs by typical lensing clusters and compared it to the RM time lag distribution of quasars across their mass range and at various redshifts. After showing that indeed galaxy clusters supply useful TDs that can easily be as large as the RM time lags expected for highest luminosity and highest redshift AGN, in the second part of this work we then estimated the number of AGN that are expected to be multiply imaged by galaxy clusters and observable with different observatories of interest. We find that for all three observatories we consider here, namely JWST, \emph{Euclid}, and \emph{Roman Space Telescope}, several hundreds to several tens of thousands of lensed AGN should be discovered by these observatories to typical depths (see Table \ref{table:TableAll}).



Before concluding the work and while perhaps quite intuitive, we now aim to provide a more visual demonstration of the idea, using an idealized case.
In particular, we consider an illustrative scenario in which the expected RM time lag is equal to the TD between two images of a quasar. 
We then discuss some of the observational challenges, that are not taken into account in the idealized example. 
We end by referring to a couple of recent lensed examples for which such RM observational campaigns may be feasible using existing facilities.

\subsection{Idealized light curve example}\label{subsec:lightcurve}
To simulate a typical continuum light curve of a quasar, we use mock light curves of the quasar HE0435-1223 generated by H0LiCOW collaboration and available online \citep{Wong2017HE0435-1223Quasar,Chen2019HE0435-1223Quasar} . We simulate the emission-line light curve by replicating the continuum light curve shifted by a certain RM time lag and add for illustration purposes, a Gaussian noise with a standard deviation $ = 0.05$ in flux. 

\begin{figure*}
    \centering
\includegraphics[width=\linewidth,trim={1.67cm 1.8cm 2.3cm 3.cm}, clip]{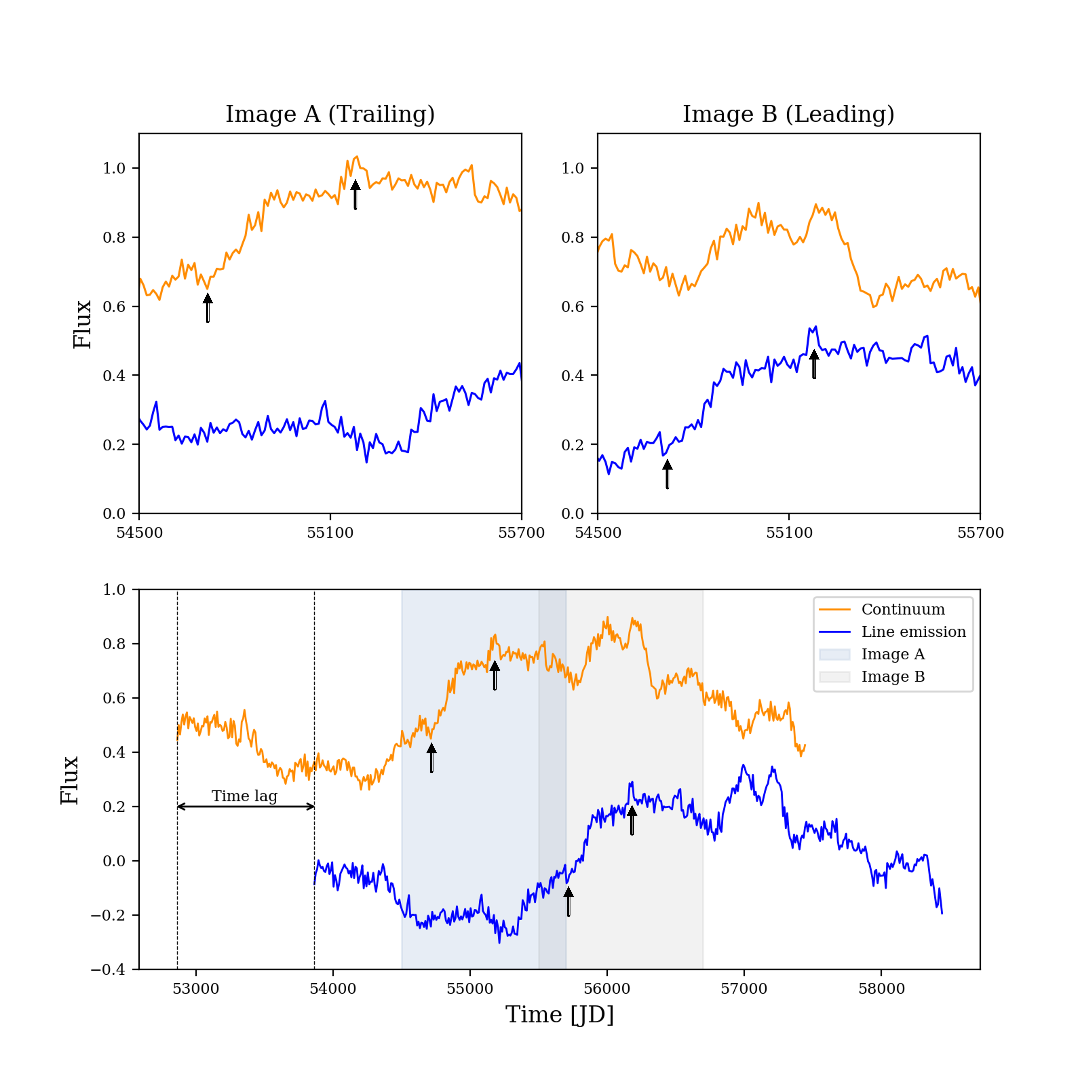}
\caption{Simulated idealized light curve for the case where the TD equals the RM time lag. In the top panel of the figure we demonstrate mock light curves of two images of the same source. The time axis represents the observing time of the images and the y axis represents a normalized flux. The SL TD between the two images is 1000 days, image A is the trailing image and image B is the leading image. The orange and blue curves correspond to the mock continuum and line emission light curves respectively. Since the two images will have different magnification factors, the leading image shown in the top right panel has its flux normalized lower, as it will usually have a lower magnification than the trailing one. Details on the mock data are given in sub-section \ref{subsec:lightcurve}. In the bottom panel we show the full mock light curves with shaded areas emphasizing the data obtained from each image and the overlap. The mock light curves of the leading image B are corrected by the TD. We emphasize the RM time lag between the continuum and emission curves at several points with arrows.}
\label{fig:lc1}
\end{figure*}

In Fig. \ref{fig:lc1} we show the mock light curves for two lensed images A and B for a SL TD of 1000 days, where it is in this case equal to the observed RM time lag. An observed BLR size of a 1000 days corresponds to a physical BLR size of 1000$/(1+z)$ light days. At a redshift of $z=2$, for example, this corresponds to a SMBH luminosity of $\lambda L_{\lambda}\sim4\times10^{47}$ erg $s^{-1}$, according to the local scaling relations. For each image we show the same time-frame of 1200 days in the observer frame, while the duration was  arbitrarily chosen for illustration purposes. For each image separately, similar to the case in which the quasar were unlensed, the pattern seen in the first 1000 days in continuum is seen repeating itself in emission lines a 1000 days later. In practice, to securely identify such an offset it would require at least several times this timeframe, hence order a decade or more of continuous monitoring. However, due to the TD being similar to the RM time lag, the same changes in the continuum seen in the trailing image are seen \emph{at the same time} in emission lines in the leading image, essentially saving the years-long wait time to observe the lagging emission line changes. Suppose that in order to detect the $\sim3$ year RM time lag a monitoring campaign spanning 15 years were needed for an unlensed source to accumulate enough data points and timespan (the true value depends on a variety of factors that we do not go into here). Investing 15 years of observing time would be looked upon as an uneasy and perhaps risky investment given that the output is uncertain. In contrast, imagine a lensed quasar such that the source is multiply imaged with at least three images, which - more often than not - would be the case for wide separation lensed quasars as we consider here (e.g.,\citealt{sharon2017,Williams2021RMlensedQSO}). In such a case the investment becomes much more reasonable: each observation triples the number of data-points compared to the unlensed case, and more importantly, assuming that $t_{TD}\sim t_{lag}$ then the corresponding emission line changes can be observed (even if not immediately identified) already in the first few batches of monitoring. Furthermore, each image would then be magnified, which would also ease the observational follow-up. 

As another example, in the case of a smaller BLR size, causing a shorter RM time lag than the SL TD, the pattern seen in the first days of observations in continuum would already be seen repeated in the emission lines after a short RM time lag. This is similar to the way it would appear for an unlensed quasar, where as for the lensed case we mainly gain magnification and additional sampling from the different multiple images, making the reverberation mapping more efficient overall. Also in the opposite scenario where the SL TD is much shorter than the expected RM time lag, while we do not gain much in terms of overall wait time, we gain in several other aspects as above. First, we obtain additional points in the light curve with the same observation, compared to the unlensed case (an advantage that increases even further when more than two quasars images are seen, as is often the case in cluster lensing). Second, the images are magnified, which facilitates their follow-up. Third, while singly imaged (or unlensed) objects can only be observed for several months a year, when they are away from the Sun's direction, a multiply-imaged quasar may allow us to fill gaps in the light curves also in those otherwise down times.
\smallskip

\subsection{Challenges and caveats}\label{subsec:obs_diff}
After demonstrating how TDs can be exploited for RM in an idealized case, we now refer to some challenges that might arise in actual RM campaigns.
First, it should be noted that throughout we assume that any object detected by a certain observatory could also be followed up spectroscopically (by the same or other observatories). In practice, follow-up may be of course more challenging, especially for those objects with larger TDs or those only reachable from space, yet in any case for all of these objects there is a gain from lensing compared to the unlensed case.

In addition, we note that even if TDs are used to facilitate RM, the monitoring time should still be sufficiently significant so as not to bias the measurements towards the available TDs. A substantial monitoring time would be needed to also account for the fact that in practice, emission-line light curves tend to be smoother compared to the continuum light curves, adding to challenges involved in assigning specific changes in the emission lines to the correct continuum changes.

We also neglect here other sources of noise such as local microlensing for some of the images, which is known to often be present and cause its own variations to the light curve for some of the components \citep[e.g.,][]{Eigenbrod2008A&A...480..647E,Mediavilla2015,Vernardos2024SSRv..220...14V} -- depending on the mass of the point mass and size of the emitting source -- although even in such cases the effects can be remedied (see e.g. \citet{Williams2021RMlensedQSO}). Similarly, we do not take into account so-called ``changing-look" AGN \citep[e.g.][]{RicciTrakhtenbrot2023NatAs...7.1282R}, for which some of the images may not show, for example, broad-line emission, while the other images might. This population of AGN seems to be extremely rare, with only about 150  discovered to date \citep{Wang2023CLAGN} and would most likely not effect our calculations. On the other hand, lensing TDs may be a route for detecting lensed AGN of this type. 

One should also bear in mind that in reality the TD is not necessarily known from the lens model to the required accuracy to remove it from the fit altogether. However, it might be first determined from the temporal shift between the continua of the two images before determining the RM time lag. This may improve the error on the TDs compared to the errors from the lens model's TD, which would otherwise propagate into the R-L relation as well.


\subsection{Future prospects}\label{subsec:prospects}
Recently, a successful reverberation mapping was performed on the multiply imaged quasar, SDSS J2222+2745, at $z=2.805$ \citep{Williams2021RMlensedQSO}, demonstrating the usefulness of multiply imaged quasars in that respect. A RM time lag of $\tau_{cen} = 36.5^{+2.9}_{-3.9}$ restframe days was significantly detected for the C\,\textsc{iv} emission line, whereas the TDs to the two other images of the quasar are about $\sim40$ and $\sim700$ days (see also \citealt{sharon2017}). Hence, one TD is of the order of the RM time lag, similar to first case discussed in \ref{subsec:lightcurve} (but with a shorter RM time lag than what we consider here, which is aimed for the high-redshift and high luminosity regimes), and the second TD is much longer than the RM time lag, similar to second case discussed in \ref{subsec:lightcurve}.

It is now worth highlighting also a couple of other, interesting recent objects which may become useful targets for RM of lensed AGN (other examples naturally exist, too). First we note the triply-imaged Little Red Dot AGN reported by \citet{Furtak2023z7p6AGN} at $=7.045$. This object is currently the highest-redshift multiply imaged AGN. It has according to their lens model a TD of $\sim$3 years, roughly, between the two closest images, and a TD 
of about 19 years between the first of these two and the third, more distant image, which is first to arrive. The NIRSpec spectrum of the source was published in \citet{Furtak2024shadows}. Based on its luminosity, with a few hours per visit we could detect both the continuum and the H$\beta$ line (no C\,\textsc{iv} lines were detected in the object, despite the very deep integration time). One can thus envision that a NIRSpec/JWST program, for example, targeting the three images simultaneously, could be used to measure both the continuum and emission lines in each epoch. From the width of H$\beta$, \citet{Furtak2024shadows} derived a BH mass of 4$\times10^7$ M$_{\odot}$, which, according to the scaling relations by \citet{Kaspi2021RM}, corresponds to an expected or observed RM time lag of a couple dozen days. This RM time lag is relatively short (corresponding to the second case discussed in \ref{subsec:lightcurve}), and while its measurement could in principle also be done were it not lensed, its measurement would be facilitated by measuring the three images simultaneously. While the expected RM time lag for this object does not appear to be very long (so that the gain in wait time is not maximized), it could supply a unique example for probing the RM relation at high-redshift.

A second, recent example is the large separation $z=3.27$ quasar COOL J0335-1927 \citep{Napier2023ApJ...954L..38NQuasarNew}, for which TDs of $\sim500$ days and $\sim130$ days between two pairs of its images were predicted from the lens model, although with quite substantial uncertainty. To estimate the expected RM time lag, we adopt the typical $g$-band luminosity seen in Fig. 4 in their work, and translate it to an absolute UV magnitude of $\sim25.5$ AB, which, based on existing scaling relations \citep{Kaspi2021RM}, corresponds to a an observed RM time lag of $\sim300$ days. The object therefore might benefit significantly from such a campaign. For more discussion on the latter object, see also \citet{Napier2023ApJ...954L..38NQuasarNew}.

We thus conclude that designated observations of multiply imaged quasars can open the door for RM studies in a much more efficient manner, as was already demonstrated at lower-redshifts \citep{Williams2021RMlensedQSO}. For the most massive and/or high-redshift quasars for which the RM time lags may be otherwise too long, this might possibly be the only, feasible way.

\section{Summary}\label{sec:Summary}
In this work we discuss the possibility of using lensing TDs to facilitate the RM of high-luminosity and high-redshift quasars, for which the radius-luminosity relation is currently, poorly constrained. The expected RM time lags for very-massive or high-redshift quasars can reach several years and would likely require much longer monitoring times than can be comfortably accommodated by typical observing campaigns. Multiply imaged AGN may, however, offer a unique opportunity to explore the radius-luminosity relation at these ends. Not only do strongly lensed quasars comprise several magnified images which boost the efficiency of the light curve sampling, but also, the TD between the quasar's multiple images can save the long wait times that are typically required. For example, if the TD is roughly of the order of the expected RM time lag, changes in the emission lines in a leading image can be observed around the same time as the changes in the continuum in a trailing image, thus allowing a much reduced wait times and overall monitoring times-spans. 

We start by showing the typical distribution of BH RM time lags compared to the lensing TD distribution from cluster lenses and show that a strong overlap exists. We then estimate the number of multiply imaged quasars that may be detectable in the near future with \textit{Euclid}, \textit{Roman}, or JWST. We find that indeed, a fair amount of multiply imaged high-redshift quasars, out of which some would experience a significantly long observed RM time lag, should be detectable. Using the TDs between the multiple images, RM of these possible future candidates could be obtained in a reasonable few-year monitoring timescale, probing the poorly known R-L relation in the early Universe.

\section*{Acknowledgements}
M.G would like to thank Yuichi Harikane for a useful discussion. M.G is thankful for the hospitality of the SURF program and cosmic DAWN center where much of this work has been carried out.
The BGU group acknowledges support by grant No. 2020750 from the United States-Israel Binational Science Foundation (BSF) and grant No. 2109066 from the United States National Science Foundation (NSF); by the Israel Science Foundation Grant No. 864/23; and by the Ministry of Science \& Technology, Israel. 
D.C. acknowledges support by grants from the German Research Foundation (HA3555-14/1, CH71-34-3) and the Israeli Science Foundation (2398/19, 1650/23).
This research made use of \texttt{Astropy},\footnote{\url{http://www.astropy.org}} a community-developed core \texttt{Python} package for Astronomy \citep{astropy13,astropy18} as well as the packages \texttt{cluster\_toolkit}, \texttt{CLASS} \citep{CLASSpython2011}, \texttt{NumPy} \citep{vanderwalt11}, \texttt{SciPy} \citep{virtanen20} and the astronomy library for \texttt{MATLAB} \citep{maat14}. The \texttt{Matplotlib} package \citep{hunter07} was used to create some of the figures in this work.

\newpage

\bibliographystyle{aasjournal}
\bibliography{MyBiblio}

\end{document}